\documentclass[twocolumn, superscriptaddress, prb, preprintnumbers, showpacs]{revtex4-1}

\usepackage{ifthen} 
\usepackage{graphicx}
\usepackage{amsmath}

\def\bra#1{\langle#1\vert}
\def\ket#1{\vert#1\rangle}
\def\braket#1#2{\langle#1\vert#2\rangle}
\def\ketbra#1#2{\lvert#1\rangle\langle#2\rvert}
\def\ketmatbra#1#2#3{\lvert#1\rangle#2\langle#3\rvert}
\def\exv#1#2#3{\langle#1\vert#2\vert#3\rangle}
\def\bfk{{\mathbf{k}}}

\begin{document}

\title{Subspace representations
in \emph{ab initio} methods for strongly correlated systems}

\author{David D. O'Regan} 
\email{ddo20@cam.ac.uk}
\affiliation{Cavendish Laboratory, University of Cambridge,
  J. J. Thomson Avenue, Cambridge CB3 0HE, United Kingdom}

\author{  Mike C. Payne}
\affiliation{Cavendish Laboratory, University of Cambridge,
  J. J. Thomson Avenue, Cambridge CB3 0HE, United Kingdom}
  
\author{Arash A. Mostofi}
\affiliation{The Thomas Young Centre, 
Imperial College London, London SW7 2AZ, United Kingdom}

\date{\today{}}

\begin{abstract}

We present a generalized 
definition of  
subspace occupancy matrices in \emph{ab initio} methods
for strongly correlated materials, such as DFT+$U$ and
DFT+DMFT, which is appropriate to the case of nonorthogonal
projector functions.
By enforcing the tensorial consistency of all 
matrix operations, we are led to a subspace projection operator
for which the occupancy matrix is tensorial
and accumulates only contributions which are
local to the correlated subspace at hand. 
For DFT+$U$ in particular,
the resulting contributions
 to the potential and ionic forces 
are automatically Hermitian, without resort to symmetrization, and
localized to their corresponding correlated subspace.
The tensorial invariance of the occupancies, energies 
and ionic forces is preserved.
We illustrate the effect
of this formalism
in a DFT+$U$ study using self-consistently determined
projectors.

\end{abstract}

\pacs{71.15.Mb, 31.15.E-, 71.15.Ap \quad
(Accepted for Physical Review B)}

\maketitle

\section{Introduction}

The routine \emph{ab initio} study of strongly correlated systems, 
that is those for which the accurate description of the physics
is beyond the capacity of mean-field methods
such as 
Kohn-Sham density functional 
theory~\cite{PhysRev.136.B864,*PhysRev.140.A1133} (DFT)
within local or semi-local  approximations to the
 exchange-correlation (XC) functional,
remains a challenge for 
electronic structure calculations. 

A number of sophisticated
methods to correct the description of strong correlation effects
within DFT have been developed which provide a good 
compromise between accuracy and computational expense.
 Successful examples include
calculations using self-interaction corrected
XC functionals~\cite{PhysRevLett.65.1148},
exact exchange in DFT~\cite{PhysRevLett.103.036404} 
 and the $GW$ approximation~\cite{PhysRevLett.74.3221}, 
 among others. 
Here we focus on methods,  
notably DFT + Hubbard $U$
(DFT+$U$)~\cite{PhysRevB.44.943, *PhysRevB.48.16929} and
DFT + dynamical mean field theory
(DFT+DMFT)~\cite{0953-8984-9-35-010,*PhysRevB.57.6884}
 for static and dynamical spatially-localized
correlation effects, respectively, which 
share a common history
and conceptual motivation based on models
for Coulomb interactions
such as the renowned Hubbard 
model~\cite{hubbard1,*hubbard2,*hubbard3}.

In such methods, the electronic system is subdivided into a 
set of spatially-localized correlated subspaces, 
the description of the interactions in which is deemed to be
beyond the capacity of the approximate XC functional, 
and the remainder which acts as a bath for particle exchange 
and for which, due to its having a large kinetic energy relative
to Coulomb repulsion, the approximate XC 
functional performs adequately.
In this manner, a model interaction may 
be used to augment and improve 
the description of the screened Coulomb interactions 
between densities in the correlated subspaces while retaining the 
computationally inexpensive XC approximation 
for the remainder
of the system. 
Generally for these methods, 
the occupancy matrix of each correlated
subspace is the object which provides the necessary
information on the electronic density to the
model describing intra-subspace interactions. 
 Defining the occupancy matrix of a correlated subspace
 using a set of orthonormal projectors is quite straightforward,
  yet the question of how to properly
 extend the formalism
 to allow for the possibility of nonorthogonal spanning functions
 is one under active debate~\cite{PhysRevB.73.045110,
 0953-8984-20-32-325205} and
 one of immediate practical consequence.
 
 It is frequently useful to permit the nonorthogonality
 of the basis functions for the 
 Kohn-Sham~\cite{PhysRev.140.A1133}
  states in \emph{ab initio} methods which make use of 
  sophisticated spatially-localized orbitals for such functions, 
  particularly in linear-scaling
 density functional theory 
 methods~\cite{PhysRevB.66.035119,PhysRevB.73.045110,
 PhysRevB.51.10157,PhysRevB.50.4316}.
 Additionally, either for reasons of computational convenience,
 as in Refs.~\onlinecite{PhysRevB.58.1201,eschrig,
 PhysRevB.73.045110}, or for the purposes of achieving 
 self-consistency over the correlated subspaces, as in 
 Ref.~\onlinecite{PhysRevB.82.081102}, it
 is common to use a  
 subset of these nonorthogonal basis functions as 
 projectors for the correlated
 subspaces, the subset termed
 \emph{Hubbard projectors}.
 We demonstrate here, however, that it may be hazardous to 
 over-identify the Hubbard projectors with the basis set from
 which they are drawn.

In this Article, we offer a revised definition of the subspace
occupancy matrix for \emph{ab initio}
methods which use nonorthogonal projectors to define the
strongly correlated subspaces. We show that, by enforcing the 
tensorial consistency
of all matrix operations, we are led immediately to a simple 
definition of the projection operator for each subspace which
is fully localized to that subspace. In contrast to previously
proposed formalisms of 
Ref.~\onlinecite{PhysRevB.73.045110}
and references therein, 
this gives rise to Hermitian corrections to the
potential and ionic forces, 
without any \emph{post hoc} symmetrization,
 which are also localized to the spaces in which the
correlation correction is required. The resulting occupancy
matrix reproduces the
 electron number of the subspaces 
 and is tensorial. Thus, for example, its trace
 is invariant under both unitary rotations
  and the generalized 
  L\"{o}wdin transformations~\cite{lowdin}
  of Ref.~\onlinecite{0953-8984-20-32-325205}.

In order to illustrate the performance of the proposed
formalism, 
we applied it to the
DFT+$U$ method in a study of two strongly
 correlated systems, 
namely bulk nickel oxide and the
gas-phase copper phthalocyanine dimer, 
with comparison to the most comprehensive 
alternative formalism available
at the time of writing, the ``dual representation'' of 
Ref.~\onlinecite{PhysRevB.73.045110}.
A set of nonorthogonal 
 generalized Wannier functions~\cite{PhysRevB.66.035119},
 optimized using the projector self-consistent DFT+$U$
 method described in Ref.~\onlinecite{PhysRevB.82.081102},
 was used in order to carry out our computational 
study with a minimum of user 
intervention in the construction of
the nonorthogonal Hubbard projectors.
 
\section{Nonorthogonal representations of 
the occupancy matrix}

Generally, in order to extract low-energy Hubbard-model
 like models from 
\emph{ab initio} DFT simulations, we require 
the projection of the single-particle density-matrix
\begin{equation}
\label{ }
\hat{\rho}^{(\sigma)} = \sum_{i \bfk}
\ket{\psi_{i \bfk}^{(\sigma)}}  f_{i \bfk}^{(\sigma)}  
\bra{\psi_{i \bfk}^{(\sigma)}},
\end{equation} 
where $\psi_{i\bfk}^{(\sigma)}$ is a Kohn-Sham eigenstate for spin
channel $\sigma$ with band index $i$, crystal momentum $\bfk$ and
occupancy $f_{i\bfk}^{(\sigma)}$,
onto a set of spatially localized subspaces. 
These subspaces 
$\mathcal{C}^{(I)}$, where $I$ is the site index, 
encompass that part of the Hilbert space of the
Kohn-Sham orbitals 
which is deemed
to be responsible for strong localized Coulomb interactions 
beyond the scope of
the approximate XC functional.

The occupancy of subspace $\mathcal{C}^{(I)}$,
which is delineated by a set of $M^{(I)}$ potentially nonorthogonal 
spanning projectors
$\ket{\varphi^{(I)}_{m}}$, $m\in\{1,\ldots,M^{(I)} \}$, 
dubbed \emph{Hubbard projectors}, 
which are associated with subspace
$I$, is generally given by the subspace-projected density matrix
\begin{equation}
\label{small_occ}
\hat{n}^{(I)(\sigma)} = \hat{P}^{(I) \dagger}  
\hat{\rho}^{(\sigma)} \hat{P}^{(I)}.
\end{equation}
The Hubbard projection operator $\hat{P}^{(I)}$, the
resolution of the identity for the space $\mathcal{C}^{(I)}$, 
is defined in terms
of the Hubbard projectors, but the exact 
manner in which this definition
should be made has been the subject of some discussion, as we 
describe in the following.

Some important conditions should be satisfied by a sound definition
 of the occupancy matrix of each correlated site, namely: 
all operations such as matrix products and traces
should be tensorially consistent so that the total
energy, potential and forces are tensorial invariants
(unaltered by arbitrary transformations of the basis on which 
the projectors for that site are defined);
any potential depending on that occupancy matrix
should be Hermitian and
its action should be strictly localized to the correlated
subspace while depending only on occupancies which are
themselves localized to that subspace; 
the trace of the occupancy matrix 
should exactly reproduce the occupancy
of the correlated manifold on that site and if the site is
extended to encompass the entire system then the
total electron number should be obtained.

\subsection{The ``full" and ``on-site" representations}

We generally assume that a set of complex, mutually nonorthogonal 
Hubbard projectors are used for each individual site and that 
the correlated subspaces possibly overlap
(we do not consider transformations 
among the projectors of different correlated sites).
Dual vectors of the Hubbard projectors must be defined
with respect to some
Hilbert superspace of the correlated manifold,
$\mathcal{H}^{(I)} \supseteq \mathcal{C}^{(I)}$, 
some possibilities for which are the subspace itself (i.e., 
$\mathcal{H}^{(I)} = \mathcal{C}^{(I)}$), 
the union of all correlated subspaces
(i.e., $\mathcal{H}^{(I)} = \bigcup_I \mathcal{C}^{(I)}$)
 and the space $\mathcal{S}$ spanned
by all basis functions in the simulation cell 
(i.e., $\mathcal{H}^{(I)} = \mathcal{S}$).
The Hubbard projector duals are 
then generally given by
\begin{equation}
\vert \varphi^{(I) m} \rangle = 
\sum_{\alpha \in \mathcal{H}^{(I)} }
\vert \varphi^{(I)}_\alpha \rangle S^{(I) \alpha m}, 
\end{equation} 
where $S^{(I) \bullet \bullet}$ is the contravariant metric tensor 
for the set of functions spanning $\mathcal{H}^{(I)}$
(the inverse of their overlap matrix). Physically meaningful 
inner products, e.g., tensorial invariants such as
occupancies, energies or forces, are computed 
between functions and elements of their
set of dual functions only (in the orthonormal case there is no 
practical distinction between functions and their duals).
For a more detailed exposition of tensor calculus applied to problems
in electronic structure theory, we refer the reader to 
Refs.~\onlinecite{Artacho,*White1997133}.

It is immediately 
clear that the simplest definition of the 
occupancy matrix for a given site, that is the projection 
$\hat{P}^{(I)} = \sum_{m \in \mathcal{C}^{(I)}}
 \ketbra{\varphi^{(I)}_{m}}{\varphi^{(I)}_{m}} $ 
of the valence manifold 
over the site's Hubbard projectors, 
\begin{equation}
\label{Eq:full}
n^{(I)(\sigma)}_{m m'} = \exv{\varphi^{(I)}_{m}}{\hat{\rho}^{(\sigma)}}
{\varphi^{(I)}_{m'}},
\end{equation}
is invalid for nonorthogonal projectors. 
This widely-used definition of the occupancy matrix, 
which is entirely appropriate in the orthonormal case,
such as calculations described in 
Ref.~\onlinecite{PhysRevB.71.035105}
and numerous citations therein,
simply neglects all nonorthogonality; 
the trace or powers
of such a fully covariant tensor
 are not physically meaningful
in the nonorthogonal case. 

A total site occupancy defined as
 a trace operation on this 
matrix, as in
\begin{equation}
N^{(I)(\sigma)} = \sum_m n^{(I)(\sigma)}_{m m},
\end{equation}
implies that such an occupancy is not, in general,
 a tensorial invariant
since it is formed by a tensorially invalid summation
over two covariant indices -- as opposed to 
a meaningful contraction of indices of opposite tensor
character. Occupancies, just like total energies, 
should be tensorial invariants,
scalars which are unchanged by 
transformations of the basis on which the 
projector functions are defined.

Progress was made in the definition of the
occupancies of correlated subspaces via 
nonorthogonal projectors
when it was noted~\cite{eschrig}
that tensorially contravariant
projector duals should be 
involved, a concept known in other contexts
for some time~\cite{mulliken}. 
A definition of the occupancy matrix fully in 
terms of Hubbard projector duals was described in 
Ref.~\onlinecite{eschrig}, for example, 
where the projection operator defined as
$\hat{P}^{(I)} = \sum_{m \in \mathcal{C}^{(I)}}
 \ketbra{\varphi^{(I) m}}{\varphi^{(I) m}} $,
provides an occupancy matrix
\begin{align}
\nonumber
n^{(I)(\sigma) m m'} ={}& \exv{\varphi^{(I) m}}{\hat{\rho}^{(\sigma)}}
{\varphi^{(I) m'}} \\ ={}&
S^{(I) m \alpha} \exv{\varphi^{(I)}_{\alpha}}{\hat{\rho}^{(\sigma)}}
{\varphi^{(I)}_{\beta}} S^{(I) \beta m'}.
\label{Eq:onsite}
\end{align}
The indices $\alpha$ and $\beta$ 
run over the spanning vectors of the
contravariant metric (i.e., the inverse overlap matrix) 
$S^{(I) \bullet \bullet}$,
on a superspace $\mathcal{H}^{(I)}$ of the 
correlated manifold $\mathcal{C}^{(I)}$.
Here and hereafter, we make use of the summation 
convention~\cite{einstein}, whereby 
repeated indices within the same expression 
are summed over unless in parentheses. 

Unfortunately, the matrix trace and powers of
Eq.~\ref{Eq:onsite} are  
not tensorially valid, as can be seen by  
taking the example of the square
of this contravariant occupancy matrix, which is of interest for 
density-density self-interaction
corrections to approximate XC functionals. The 
resulting expression for the squared occupancy matrix 
\begin{equation}
n^{2 (I)(\sigma) m m'} =  \sum_{m'' \in \mathcal{C}^{(I)}} 
n^{(I)(\sigma) m m''}  n^{(I)(\sigma) m'' m'} 
\end{equation}
implies that the operator 
\begin{equation}
\hat{P}^{(I)} = \sum_{m'' \in \mathcal{C}^{(I)}} 
 \ketbra{\varphi^{(I) m''}}{\varphi^{(I) m''}} 
\end{equation}
forms a tensorially traceable identity on $\mathcal{C}^{(I)}$. 
It does not in the case of
nonorthogonal projectors, however, 
since an identity operator may only 
be formed via the outer product between a projector and a 
projector dual, and not a dual vector and its own complex
conjugate, 
including the case 
where the correlated subspace
is extended to the Hilbert space of the 
entire system ($\mathcal{C}^{(I)} = \mathcal{S}$).

The shortcomings in the two definitions of the occupancy matrix 
described above have been 
previously described in detail by Han et. al. in 
Ref.~\onlinecite{PhysRevB.73.045110} and are
dubbed, respectively, 
the ``full'' (Eq.~\ref{Eq:full}) and 
``on-site'' (Eq.~\ref{Eq:onsite}) representations in the nomenclature 
described therein. 
These authors concentrated on the special
case where the dual-generating superspace $\mathcal{H}^{(I)}$ 
is the space spanned by all basis
functions $\lbrace \lvert \phi_\alpha \rangle \rbrace$ 
in the simulation cell, so that $\mathcal{H}^{(I)} = \mathcal{S}$,
in which case 
$S_{\alpha \beta} = \braket{\phi_\alpha}{\phi_\beta}$
and the Hubbard projectors form a subset of the basis set.  
Thus, the same contravariant metric for all basis
functions in the simulation cell is used to generate the dual functions
on each correlated site and in this case
the ``full'' and ``on-site'' occupancy matrices simplify,
respectively, to 
\begin{equation}
n^{(I)(\sigma)}_{m m'} = \sum_{\alpha, \beta \in \mathcal{S}}
S_{m  \in \mathcal{C}^{(I)} \alpha} K^{(\sigma) \alpha \beta} 
S_{\beta m'  \in \mathcal{C}^{(I)} } 
\end{equation}
and 
\begin{equation}
n^{(I)(\sigma) m m'} =  K^{(\sigma) m \in \mathcal{C}^{(I)} m'  \in \mathcal{C}^{(I)} }.
\end{equation} 
Here, $ K^{(\sigma) \alpha \beta} = 
\exv{\phi^\alpha}{\hat{\rho}^{(\sigma)}}
{\phi^\beta}$ is the representation of the density matrix in terms 
of basis-set duals, known as the density kernel. 
The notation $S_{m  \in \mathcal{C}^{(I)} \alpha}$
 reminds us that $m$ and $\alpha$ run over the spanning vectors
 of two different spaces, $\mathcal{C}^{(I)}$ and 
 $\mathcal{H}^{(I)} = \mathcal{S}$, respectively, so that the block of 
 $S_{\bullet \bullet}$ in question is generally not square.

\subsection{The ``dual" representation}

Han et. al.~\cite{PhysRevB.73.045110}, whose 
invaluable contribution on this
subject addressed many of the salient issues,
pointed out that the total number of electrons
is not recovered by the trace of the occupancy matrix 
if the site is extended to 
include the entire simulation cell 
using the ``full'' and ``on-site'' representations.
They proposed an alternative ``dual''
representation which solves this particular problem
and is generated by the projector
\begin{equation}
\hat{P}^{(I)} = \frac{1}{2 } 
\sum_{m \in \mathcal{C}^{(I)}} \left(
 \ketbra{\varphi^{(I) m}}{\varphi^{(I)}_{m}} +  
 \ketbra{\varphi^{(I)}_{m}}{\varphi^{(I) m}}
 \right)
 \end{equation}
 and the corresponding occupancy matrix
 \begin{align} \nonumber
{}& 
\frac{1}{2} \left( 
\exv{\varphi^{(I) m}}{\hat{\rho}^{(\sigma)}} {\varphi^{(I)}_{m'}} 
+ 
\exv{\varphi^{(I)}_{m}}{\hat{\rho}^{(\sigma)}} {\varphi^{(I) m'}} 
\right) \\
={}&
\frac{1}{2}  \sum_{\alpha \in \mathcal{S}} \left( 
\begin{array}{cc}
K^{(\sigma) m  \in \mathcal{C}^{(I)} \alpha } 
S_{\alpha m'  \in \mathcal{C}^{(I)} }  + \\
S_{m  \in \mathcal{C}^{(I)} \alpha} K^{(\sigma) 
\alpha m'  \in \mathcal{C}^{(I)} } 
 \end{array}  \right).
\end{align}

In the ``dual" representation, the contravariant metric on the
basis set is used to form the Hubbard projector duals 
(which are therefore 
delocalized across the entire simulation cell, in general,
 since the inverse overlap
matrix is dense even when the overlap matrix itself
is sparse)
via $\lvert {\varphi^{(I) m}} \rangle = 
\sum_{\alpha \in \mathcal{S}} \lvert {\varphi^{(I)}_{\alpha}} \rangle 
S^{ \alpha m}$. 
Symmetrization is then carried out 
in order to both provide
a symmetric occupancy matrix and to 
recover a Hermitian potential.

The ``dual'' representation shares with the ``full'' 
representation the 
attribute of Hermiticity  
and, furthermore, it 
has a tensorially and physically meaningful trace.
As such, to our knowledge, 
it provides the most favourable 
occupancy definition hitherto available.
However, this occupancy matrix is tensorially 
ambiguous, consisting of the sum of 
tensors of differing index character.

One cannot generally symmetrize or antisymmetrize 
a tensor over indices of mixed
covariant and contravariant character in this 
way and obtain a matrix
which transforms as a tensor 
(one which may be used to generate tensorially invariant
occupancies, local moments or energies).
Thus, while providing a significant improvement over previously
suggested definitions of the 
occupancy matrix due to its tensorially
invariant trace, the ``dual'' representation
 suffers similar problems with matrix powers as other
 representations: if we attempt to 
 compute the square of this matrix 
 we obtain tensorially inconsistent,
  and thus physically meaningless,
 terms in the product of the form $n^{m}_{\;\;m''}
 n_{m''}^{\;\;m'} $ and 
 $n_{m}^{\;\;m''} n^{m''}_{\;\;m'} $.
 
 \subsection{Requirement for a subspace-localized 
 Hermitian projection operator}
 
 Let us step back for a moment and consider why 
 any projection operator of the form
\begin{align} 
\hat{P}^{(I)} {}&= 
\sum_{\substack{m \in \mathcal{C}^{(I)} \\ \alpha \in 
\mathcal{H}^{(I)} \ne \mathcal{C}^{(I)} } }
\ketmatbra{\varphi^{(I)}_{ \alpha}}{S^{\alpha m   } }
{\varphi^{(I)}_{ m}} \end{align} requires symmetrization to the ``dual''
form in order to provide a Hermitian potential operator. 

An arbitrary potential
operator $\hat{V}$, operating on the 
subspace $\mathcal{C}^{(I)}$, 
which could represent the screened Coulomb interaction, 
for example,
has matrix elements in the frame of Hubbard projectors given by
\begin{align}
V^{(I) m'}_m 
=
\sum_{\alpha \in \mathcal{H}^{(I)}} 
 \exv{\varphi^{(I)}_{m}}{\hat{V}}{\varphi^{(I)}_\alpha} S^{\alpha m'}.
\end{align}
The potential operator is easily 
shown to be non-Hermitian in the case
where $m, m' \in \mathcal{C}^{(I)} \subset \mathcal{H}^{(I)}
\subseteq \mathcal{S}$, and 
$\mathcal{C}^{(I)} \ne \mathcal{H}^{(I)}$ strictly holds, since
\begin{align} 
\hat{V}^{(I)} ={}& \hat{P}^{(I) \dagger} \hat{V} \hat{P}^{(I)} =
\ketmatbra{\varphi^{(I)}_m}{V^{(I) m m'}}{\varphi^{(I)}_{m'}} \\
\nonumber
={}&  \sum_{\alpha, \beta \in \mathcal{H}^{(I)}}
\ket{\varphi^{(I)}_m} S^{(I) m \alpha} V^{(I)}_{\alpha \beta} 
S^{(I) \beta m'}
\bra{\varphi^{(I)}_{m'}} \\ \nonumber
\neq{}&  \sum_{\alpha, \beta \in \mathcal{H}^{(I)}}
\ket{\varphi^{(I)}_{\alpha}} S^{(I) \alpha m} V^{(I)}_{m m'} S^{(I) m' \beta}
\bra{\varphi^{(I)}_{\beta}} \\
={}& 
\ketmatbra{\varphi^{(I) m}}{V^{(I)}_{ m m'}}{\varphi^{(I) m'}} =
\hat{P}^{(I) } \hat{V} \hat{P}^{(I) \dagger} = \hat{V}^{(I)\dagger} . \nonumber
\end{align}

The reason for this non-Hermiticity is that the indices 
$\alpha, \beta $ do not generally
run over functions spanning just the correlated 
space $\mathcal{C}^{(I)}$, but rather over those that span a 
superspace $\mathcal{H}^{(I)}$, e.g., 
typically over the basis functions in the simulation cell,
$\mathcal{H}^{(I)} = \mathcal{S}$. 
This observation is quite general:
the dual projectors must be constructed using the 
metric on \emph{precisely} the space spanned by
the projectors themselves in order to build a Hermitian
projection operator and hence a Hermitian
potential. This cannot be circumvented in a tensorially-consistent way 
by symmetrizing operators since tensors can only be symmetrized
over pairs of indices if they are either both of covariant character or both
of contravariant character.

\subsection{The ``tensorial" representation}

Based on the above, we deduce 
that, in order to build a 
tensorially-consistent occupancy matrix
which generates a Hermitian potential, 
the projection operator for a given 
subspace $\mathcal{C}^{(I)}$ 
must necessarily be constructed using 
exact dual Hubbard projectors with 
respect to that subspace only. 
Thus, with the covariant 
overlap matrix of Hubbard 
projectors defined by
\begin{equation}
O^{\left( I \right)}_{m m' } =
\langle \varphi ^{\left( I \right) }_m \rvert
\varphi ^{\left( I \right) }_{m'} \rangle,
\end{equation}
that is an \emph{individual} 
$M^{(I)} \times M^{(I)}$ covariant metric tensor for 
each correlated site $I$,
the proper dual vectors $ \ket{\varphi^{\left( I \right) m}} $ 
are constructed
using the corresponding contravariant metric $O^{\left( I \right) m' m }$,
 as per
\begin{equation}
\lvert \varphi^{\left( I \right) m} \rangle = 
 \sum_{m' \in \mathcal{C}^{(I)}}
\lvert  \varphi^{\left( I \right)
}_{m'} \rangle O^{\left( I \right) m' m }.
\end{equation}
Here, we emphasize that the contravariant metric is obtained via
a separate $M^{(I)} \times M^{(I)}$ inverse operation for each site,
so that
\begin{equation}
O^{\left( I
\right) m' m'' } O^{\left( I \right)}_{m'' m } = \delta^{m'}_{\;\; m}.
\end{equation}

In the special case where the Hubbard
 projectors are drawn from the set of functions used to
 represent the Kohn-Sham 
 wave-functions, the 
 overlap matrix of duals $O^{(I) \bullet \bullet}$ 
 for each site cannot generally
 be extracted immediately from the metric $S^{\bullet \bullet}$ 
 on $\mathcal{S}$.
However, in this particular case, the $O^{(I)}_{\bullet \bullet}$ 
matrix for each site is merely a sub-block of the 
basis-function overlap $S_{\bullet \bullet}$ and, from 
this, the contravariant $O^{(I) \bullet \bullet}$ 
for each site can be computed
by a separate inverse operation for each site
which is typically fast, due to the small matrix dimension.

Employing this definition of the metric tensor on each subspace, 
the projector duals remain
manifestly as localized to the correlated 
subspace as the projectors themselves, they pick up only 
subspace-localized contributions to the occupancy and can only
apply subspace-localized corrective potentials, as we would expect
for local self-interaction corrections such as DFT+$U$ or its extensions. 
The Hubbard projection operator, in what we will denote the
``tensorial'' representation,
 \begin{equation} \hat{P}^{(I)} =
\sum_{m, m' \in \mathcal{C}^{(I)} }  
\ketmatbra{\varphi^{(I)}_{ m'}}{O^{m' m   } }
{\varphi^{(I)}_{ m}} \label{eq:projection} \end{equation}
is Hermitian and thus gives rise to a Hermitian potential, 
without resort to symmetrization, since
$O^{(I) \bullet \bullet}$ is a square overlap matrix, 
\begin{align} 
\hat{V}^{(I)} ={}& \hat{P}^{(I) \dagger} \hat{V} \hat{P}^{(I)} =
\ketmatbra{\varphi^{(I)}_{ m}}{V^{(I) m m'}}{\varphi^{(I)}_{m'}} \\
={}& \nonumber
\ket{\varphi^{(I)}_m} O^{(I) m m'} V^{(I)}_{m' m''} O^{(I) m'' m'''}
\bra{\varphi^{(I)}_{m'''}} \\
={}& 
\ketmatbra{\varphi^{(I)  m}}{V^{(I)}_{ m m'}}{\varphi^{(I) m'}} =
\hat{P}^{(I) } \hat{V} \hat{P}^{(I) \dagger} = \hat{V}^{(I)\dagger}.
\nonumber
\end{align}

The occupancy matrix is most easily expressed in its singly covariant
and singly contravariant form, though other forms are readily obtainable 
from the metric tensor, so manipulations of the following form can be made:
\begin{equation}
n_\bullet^{\;\; \bullet} = O_{\bullet \bullet} n^{\bullet \bullet}
=  n_{\bullet \bullet} O^{\bullet \bullet} = 
O_{\bullet \bullet} n^\bullet_{\;\; \bullet} O^{\bullet \bullet}.
\end{equation}
The contravariant-covariant or covariant-contravariant forms of the 
tensorial occupancy matrix, the latter given by 
(the second line applies to the special case where 
Hubbard projectors are drawn
from the basis set)
\begin{align}
n_m^{(I)(\sigma) m'} ={}& \nonumber
 \exv{\varphi^{(I)}_{m}}{\hat{\rho}^{(\sigma)}}
{\varphi^{(I)}_{m''}} O^{(I) m'' m'} \\
={}& \sum_{\alpha, \beta \in \mathcal{S} }
S_{m  \alpha}
K^{\alpha \beta} S_{\beta m''}
O^{(I) m'' m'},  \end{align}
possess a common tensorially invariant trace
(a proper contraction over one covariant and one contravariant
index) 
which recovers the exact number of electrons in the correlated subspace
by construction (the so-called sum-rule), so that
\begin{equation}
N^{(I)(\sigma)} =    \sum_{m \in \mathcal{C}^{(I)} }
n^{(I)(\sigma)m}_{\;\;\;\;\;\;\; m} = 
 \sum_{m \in \mathcal{C}^{(I)} }
n^{(I)(\sigma)m}_{\;\;\;\;\;\;\;\;\;\;\; m} .
\end{equation}

Their powers themselves remain  tensors, for example 
the square
$n_{\;\;\;\;\;\;\;\; m}^{2 (I)(\sigma) m'} = n_{\;\;\;\;\;\;\; m}^{(I)(\sigma) m''} 
n_{\;\;\;\;\;\; m''}^{(I)(\sigma) m'} $ 
is itself a well behaved singly covariant and singly contravariant tensor
with an invariant trace.
This occupancy matrix trace is easily
demonstrated to be invariant under rotations of the set of Hubbard
projectors on its site and it is independent of the basis used to represent
the Kohn-Sham states.  

An occupancy matrix which is
invariant under element-wise transpose
might lend itself to an interpretation as quantifying
the charge shared between Hubbard projectors, 
and indeed it does in the case
of orthonormal Hubbard projectors. However, it is worth emphasizing
that in the case of a set of nonorthogonal Hubbard projectors, 
these functions
are merely spanning vectors with no rigorous
physical meaning and, generally, no such interpretation of charge
shared between orbitals may be safely made.
In fact, in the nonorthogonal case, the
occupancy matrix should not be generally expected to be 
invariant under element-wise transpose, i.e.,
$n_m^{\;\; m'} \ne n_{m}^{T m'} =  n^{m'}_{\;\; m}$. 
Rather, if the duals are defined in a way which preserves the
 tensorial consistency of inner products, the occupancy matrix 
 must satisfy instead 
 the more general expression 
 $n_m^{\;\; m'} = O_{m m''} n^{m''}_{\;\;\;m'''} O^{m''' m'}$, where
$O_{\bullet \bullet}$ and $O^{\bullet \bullet}$ are the covariant and 
contravariant metric tensors, respectively, on the subspace in question.
As a result, only the diagonal elements of the occupancy matrix
can be imbued with a intuitive
 meaning in the sense of occupancy; 
symmetrizing the matrix does not 
recover such an interpretation for the
off-diagonal elements.
  
\section{Application to the DFT+$U$ method}

In this section, we illustrate the practical application of the
``tensorial" representation to a particular method 
for strongly correlated materials, namely the simplified 
rotationally invariant DFT+$U$ correction of 
Refs.~\onlinecite{PhysRevB.71.035105,PhysRevLett.97.103001}.
We provide the necessary expressions for the
tensorially invariant
DFT+$U$ terms in the energy, potential and ionic forces
for use with nonorthogonal Hubbard projectors,
which is of some importance since 
such a set is often used in
contemporary high-accuracy, particularly linear-scaling, 
implementations~\cite{PhysRevB.66.035119,PhysRevB.73.045110,
 PhysRevB.51.10157,PhysRevB.50.4316}.
We have recently shown that
an efficient set of Hubbard
projectors can be constructed which is 
self-consistent with the set of
truncated nonorthogonal generalized Wannier functions
which minimize the 
DFT+$U$ total energy~\cite{PhysRevB.82.081102}.
 
In the DFT+$U$ approach, 
a set $M^{(I)}$ Hubbard projectors,
typically spatially localized
on a particular transition-metal or Lanthanoid
atom, is used to
define the occupancy matrix of the correlated
 subspace at each site.
The particles occupying these  
subspaces interact
strongly with each other by comparison 
with their interaction with the bath; each subspace acts as
an open quantum system. As such, 
we may separately impose the Fock antisymmetry condition
for the projected wave-function for
each strongly correlated subspace, so that the subspace occupancy, 
the projection of the full single-particle density
onto the subspace in question, should itself
be a valid density-matrix operator (it should be idempotent
and reproduce the electron number of that subspace). 

Since the idempotency of the density matrix for the
full system is a condition which must be exactly satisfied, 
and the idempotency of correlated
sites is a competing condition 
(the Hubbard projectors differ from Kohn-Sham orbitals in general), 
the subspace idempotency may be only partially enforced,
for each correlated site, using an idempotency 
penalty functional 
of the form
\begin{equation}
\sum_{I \sigma} Tr 
\left[ \hat{\lambda}^{( I) (\sigma)} \left(
\hat{n}^{\left( I \right) \left( \sigma \right)} - \hat{n}^{\left( I
\right) \left( \sigma \right) 2} \right) \right],
\end{equation}
which disfavours deviation from 
wave-function anti-symmetry 
in the strongly correlated subspaces.
The pre-multiplier $\hat{\lambda}^{( I) (\sigma)}$
is usually approximated by a
 single scalar for each site, where it is identified as
\begin{equation}
 \hat{\lambda}^{\left( I \right)\left( \sigma \right)} = 
 \frac{ {U}^{\left( I \right)\left( \sigma \right)}}{2},
\end{equation}
half of the  
screened, subspace-averaged 
Coulomb  
interaction. If we further assume an 
orthonormal set of Hubbard projectors
for each site, the functional is easily recognisable as
the familiar rotationally invariant  
DFT+$U$ correction term of Cococcioni and
de Gironcoli in Ref.~\onlinecite{PhysRevB.71.035105}, 
\begin{equation}
 \sum_{I \sigma} \frac{ {U}^{\left( I \right)\left( \sigma \right)}}{2} 
  {\left[ \sum_{m}
 n_{m m} - \sum_{m m'} n_{m m'}
  n_{m' m}\right]}^{(I)(\sigma)}.
\label{eq:EU}
\end{equation}

\subsection{The tensorially invariant DFT+$U$ functional}

Let us consider how we might generalise this 
DFT+$U$ penalty-functional to accommodate an 
orbital-dependent interaction tensor. 
The Coulomb interaction tensor $U$ for a given spin channel
and site (considering the same Hubbard projectors for different
spins for brevity of notation)
is given generally by the two-centre integral
(N.B. using the Dirac, and not Mulliken, convention)
\begin{equation}
U^{(I)}_{m m' m'' m'''} = 
\langle \varphi^{(I)}_{m\phantom{'}}  \varphi^{(I)}_{m'} 
\rvert \hat{U}^{(I)(\sigma)} \left( \mathbf{r}, \mathbf{r'} \right) 
\lvert \varphi^{(I)}_{m''} \varphi^{(I)}_{m'''}  \rangle .
\label{eq:Umatrix}
\end{equation} 
Here, $\hat{U}^{(\sigma)} \left( \mathbf{r}, \mathbf{r'} \right) $ 
is the Coulomb interaction screened according to 
mechanisms described by an appropriate theory
such as 
linear-response~\cite{PhysRevB.58.1201,
PhysRevB.71.035105,PhysRevLett.97.103001}, 
constrained LDA~\cite{PhysRevB.39.1708,*PhysRevB.43.7570,
*PhysRevB.74.235113}, 
constrained RPA~\cite{PhysRevB.70.195104,*PhysRevB.81.245113} or 
constrained adiabatic LDA~\cite{PhysRevB.74.125106}.
Coulomb repulsion is represented by those terms for which
$m=m'';m'=m'''$, while direct exchange is given by those 
elements with $m=m''';m'=m''$.

In the general, nonorthogonal case, care must be
taken in employing the $U$ tensor
in order to preserve the tensorial 
invariance of the DFT+$U$ energy. 
For example, if a tensorial invariant is required which 
provides
the sum of the part of the tensor describing density-density
Coulomb repulsions, it should correctly be computed by contracting 
covariant and contravariant indices in pairs of indices of
opposite character, i.e.,
double-sums of the form, where $m, m' \in\{1,\ldots,M^{(I)} \}$,
\begin{equation}
U_{m m'}^{\quad m m'}, \quad U^{m m'}_{\quad m m'}, \quad 
U_{m \;\;\;\;\;\; m'}^{\;\; m' m}, \quad  
\mbox{or} \quad U^{m \;\;\;\;\;\; m'}_{\;\;\; m' m}
\end{equation}
are admissible while those of the form
$U_{m m' m m'}$ or $U^{m m' m m'} $ 
break tensorial invariance.
Indices are raised and lowered simply using the metric tensor
of the correlated subspace to which the
$U$ tensor corresponds, the contravariant $O^{\bullet \bullet}$ 
or covariant $O_{\bullet \bullet}$, respectively, e.g.,
\begin{equation}
U_{m \;\; m'}^{\;\; m' \;\; m} = O^{m' m''} U_{m m'' m' m'''}  O^{m''' m} . 
\end{equation}
Purely as an illustration of this principle,
a simple projector-decomposed 
tensorially invariant penalty functional 
may be constructed using  
pairwise contractions over the four indices, as in 
\begin{equation}
 \sum_{I \sigma}  \frac{1}{2}  
 U_{m m''}^{(I) \;\; m' m'''}  \left[ 
 n_{m'}^{\;\;\; m} \delta_{m'''}^{\;\; m''}
-  n_{m'}^{\;\;\; m''}
  n_{m'''}^{\;\;\; m}\right]^{(I)(\sigma)}.
\end{equation}

A commonly used approximation for the
screened Coulomb interaction, 
that which we use, is where
the interaction tensor (itself an inverse response function)
is averaged over the subspace (i.e., both over perturbing 
and probing indices),
providing a scalar density-density 
Coulomb interaction.
The usual DFT+$U$ penalty functional in this fully averaged
approximation is thus given, in tensorially-invariant form, by
the expression 
\begin{equation}
 \sum_{I \sigma}  \frac{1}{2 M^{(I)2}} 
 U_{m'' m'''}^{(I) \;\;\; m'' m'''}  \left[ 
 n_{m}^{\;\; m} 
-  n_{m}^{\;\; m'}
  n_{m'}^{\;\; m}\right]^{(I)(\sigma)}.
\end{equation}

\subsection{DFT+$U$ potential and ionic forces
in the tensorial formalism}

The DFT+$U$ term in the Kohn-Sham potential,
generally given (for real valued $U$ tensors) by
\begin{equation}
 \hat{V}^{(\sigma)} = \sum_{I }   
\lvert \varphi^{(I) m}\rangle V_m^{(I)(\sigma) m'} 
 \langle \varphi^{(I)}_{m'}\rvert ,
\end{equation}
has matrix elements, in the case of averaged $U$, given by
\begin{equation}
 V_m^{(I)(\sigma) m'} =  \frac{1}{2 M^{(I)2}}  
  U_{m'' m'''}^{(I) \;\;\; m'' m'''}
 \left[  \delta_{m}^{\;\; m'}   - 2 n_{m}^{(I)(\sigma) m'}
  \right]. \nonumber
\end{equation}
The DFT+$U$ potential 
is Hermitian by construction when the Hubbard
projection operator built with the subspace-local
tensorial representation of Eq.~\ref{eq:projection}, is used.
No symmetrization of the occupancy matrices is then needed
to ensure this Hermiticity
and the potential acts strictly within the spatial extent of the
subspace of whose occupancy it depends.

Correspondingly, the DFT+$U$ contribution to the 
force on the ion labelled $J$, with position
$\mathbf{R}_J$, is given by
\begin{align} \nonumber
 \mathbf{F}_{J} = -{}& \sum_{I \sigma} 
  \Big\langle \frac{d \varphi^{(I)}_m}{d \mathbf{R}_J}
  \Big\rvert \varphi^{(I)}_{m'} \Big\rangle O^{(I) m' m''} 
  n_{m''}^{(I)(\sigma) m'''}
  V_{m'''}^{(I)(\sigma) m} \\ \nonumber
  -{}& \sum_{I \sigma} 
 n_{m}^{(I)(\sigma) m'}
  \Big\langle  \varphi^{(I)}_{m'} 
  \Big\rvert \frac{d \varphi^{(I)}_{m''}}{d \mathbf{R}_J} \Big\rangle 
  O^{(I) m'' m'''} V_{m'''}^{(I)(\sigma) m}.
\end{align}
Here, we have made simplifications such as
\begin{equation}
\Big\lvert \frac{d \varphi^{(I) m}}{d \mathbf{R}_J} \Big\rangle =
\Big\lvert \frac{d \varphi^{(I)}_{m'}}{d \mathbf{R}_J} 
\Big\rangle  O^{(I) m' m} ,
\end{equation}
which is valid if the 
subspace metric tensor is position independent,
in particular if the 
Hubbard projectors are simply
spatially translated when their host ion is moved.
Our force expression holds, of course, only if we are on the 
Hellmann-Feynman surface, where the density-matrix 
commutes with the Hamiltonian. 

\section{Bulk Nickel Oxide}

The first row transition-metal monoxide NiO 
poses some difficulties to 
Kohn-Sham density-functional theory 
and to electronic structure theories
generally. As such, it has served as a valuable 
proving-ground for approaches such as
periodic unrestricted Hartree-Fock 
theory~\cite{PhysRevB.50.5041}, the self-interaction corrected
local density approximation \cite{PhysRevLett.65.1148},
the $GW$ approximation \cite{PhysRevLett.74.3221} and
DFT+DMFT \cite{PhysRevB.74.195114}. 
Experimentally, the paramagnetic phase of NiO 
is found to possess a rock-salt crystal 
structure with 
a lattice constant of approximately 
$4.17 $~\AA~\cite{PhysRevB.57.1505}. 
At ambient temperature,
NiO is a type-{\small {\sc II}} antiferromagnetic 
insulator with a local 
magnetic moment of between 
$1.64~\mu_B$ and $1.9~\mu_B$~\cite{PhysRevB.71.035105}.
Due to the persistence of its magnetic moment 
and optical gap, which is approximately $4~$eV,
above the N\'{e}el temperature, 
it falls broadly into the 
category of a Mott insulator~\cite{PhysRevB.50.5041}
with a charge-transfer insulating gap
of predominantly oxygen $2p$ to
nickel $3d$-orbital 
character~\cite{PhysRevB.50.5041,PhysRevLett.53.2339}. 
It has long been recognised that LDA-type exchange
correlation functionals~\cite{PhysRevB.23.5048} 
qualitatively fail to reproduce the 
physics of this material, grossly under-estimating the local
magnetic moment, the Kohn-Sham gap and assigning an
incorrect fully nickel $3d$-orbital character 
to the valence band edge. We stress, however, that the
Kohn-Sham gap is not comparable to the experimental insulating
excitation gap, even for the exact XC 
functional~\cite{PhysRevLett.51.1884}.

The DFT+$U$ method has previously been applied, in numerous 
incarnations, to bulk NiO and 
it is known to recover the 
 principal features of this strongly-correlated
 oxide~\cite{PhysRevB.44.943, *PhysRevB.48.16929, 
PhysRevB.57.1505, PhysRevB.62.16392, PhysRevB.58.1201, 
PhysRevB.71.035105}.
Moreover, generalizations to DFT+$U$ such as first-principles 
methods for calculating the Hubbard $U$ 
parameter~\cite{PhysRevB.58.1201,PhysRevB.71.035105},
the DFT+$U$+$V$ method for including inter-site 
interactions~\cite{0953-8984-22-5-055602} and,
most pertinent for this study, previous investigations into
 subspace representations of nonorthogonal Hubbard
 projectors in DFT+$U$~\cite{PhysRevB.73.045110,
 0953-8984-20-32-325205} have also been applied
 successfully to this system.
 We have chosen to study NiO, therefore, because it
 is so well characterized and 
 we have performed calculations
 which we hope will be complementary to those described in 
 Ref.~\onlinecite{PhysRevB.73.045110}, where the 
 ``full", ``on-site" and ``dual" representations of a 
 linear-combination of pseudo-atomic orbital basis were 
 compared.
   
\subsection{Computational methodology}

Calculations of the 
ground-state electronic
structure of bulk antiferromagnetic nickel oxide 
were carried out within collinear spin-polarized Kohn-Sham 
DFT~\cite{PhysRev.136.B864,PhysRev.140.A1133},
 and the simplified DFT+$U$ method~\cite{PhysRevB.71.035105}. 
 The linear-scaling \textsc{ONETEP} first-principles 
package, described in detail in 
Refs.~\onlinecite{onetep1,*ChemPhysLett.422.345}, was used.
The LSDA (PZ81) exchange-correlation 
functional~\cite{PhysRevB.23.5048}, with norm-conserving
pseudopotentials~\cite{PhysRevB.41.1227,opium},
was invoked throughout.
Periodic boundary conditions were used with
a $512$-atom supercell  
and the Brillouin zone was sampled at the $\Gamma$-point only.
A systematic variational basis of Fourier-Lagrange,
also known as periodic cardinal sine or \textit{psinc}, 
functions~\cite{mostofi-jcp03,*Baye1986}, 
was used, 
equivalent to a set of plane-waves
bandwidth limited to a kinetic-energy 
cutoff of $825~$eV. 

\begin{figure}
\includegraphics*[width=8.5cm]
{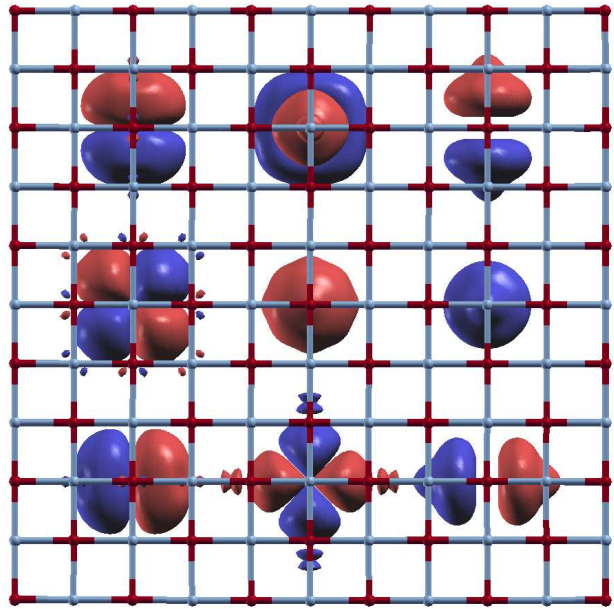}
\caption{(Color online) Isosurfaces of the set of
nonorthogonal generalized
Wannier functions (NGWFs) on a 
nickel atom in NiO.
The NGWFs are those computed at projector
self-consistency in the ``tensorial'' representation
at LDA+$U=6$~eV. Those in the left column
(predominantly $3d-t_{2g}$ character) 
and the top and bottom NGWFs
in the middle column 
(predominantly $3d-e_g$ character)
are those used as Hubbard projectors, 
while the remaining NGWFs
(pseudized $4s$-like in the centre and 
pseudized $4p$-like in the right column) lie 
outside the correlated subspace on that atom.
The isosurface is set to half of the maximum for the 
$4s$ and $4p$-like NGWFs and $10^{-3}$ times the
maximum for the $3d$-like NGWFs.}
\label{Fig:NiO_SCF1_Ueq6_FINALNGWFs}
\end{figure}

In the \textsc{ONETEP} method,
 the Kohn-Sham density-matrix is represented
in the separable form
\begin{equation}
\rho^{(\sigma)} \left( \mathbf{r} ,  \mathbf{r'}\right)
= \phi_\alpha \left( \mathbf{r} \right)
K^{(\sigma) \alpha \beta}
\phi_\beta \left( \mathbf{r'} \right)
\end{equation}
 in terms of a set of covariant
nonorthogonal generalized 
Wannier functions (NGWFs)~\cite{PhysRevB.66.035119},
$\lbrace \phi_\bullet \left( \mathbf{r} \right) \rbrace$,
and a corresponding contravariant density kernel,
$K^{\bullet \bullet }$, for each spin channel.
The density kernel was not truncated in the
calculations described here.
In the  \textsc{ONETEP} 
method~\cite{onetep1,*ChemPhysLett.422.345}, the
 total energy is iteratively minimized 
both with respect to the elements
of the density kernel
for a given set of NGWFs~\cite{0953-8984-20-29-294207}, 
using a combination of 
penalty-functional~\cite{RevModPhys.32.335} and 
LNV~\cite{PhysRevB.47.10891,
*PhysRevB.50.17611,*PhysRevB.47.10895}
techniques to ensure the validity 
of the density matrix, and with respect to the expansion
coefficients of the NGWFs in the psinc basis.
The converged NGWFs (a minimal set of 
nine functions for nickel $4s$, $4p$ and $3d$ and
four for oxygen $2s$ and $2p$, truncated to an atom-centered
sphere of $4.0~$\AA, were employed
 in calculations on NiO) are those 
which are optimized to minimize the total energy and are thus
adapted to the chemical environment, incorporating all
valence-electron
 hybridization effects in the ground-state density.

Our principal purpose was to provide an appraisal 
of the difference in predicted electronic 
properties, if any, given by 
DFT+$U$ when using nonorthogonal Hubbard projectors
with either the ``dual'' or ``tensorial'' 
representations of the
correlated subspaces.
The ``dual'' representation, in particular,
 was selected for comparison since it appears to be the 
 most sophisticated of the previously proposed 
 subspace definitions -- 
it has a tensorially invariant
 occupancy matrix trace which cannot be 
 said of the manifestly incomplete ``on-site" and
 ``full" representations. The latter three representations
 were previously compared in detail in 
 Ref.~\onlinecite{PhysRevB.73.045110}. 

The underestimation 
of the NiO lattice parameter with respect to experiment
by the LDA,
as well as the ability of DFT+$U$ to correct this, for a
particular $U$ value, has been known for some 
time~\cite{PhysRevB.57.1505}. 
The $U$ parameter required to correct 
the lattice will depend on details 
of the underlying XC functional, the precise DFT+$U$
functional used, the pseudopotentials and
both the form of the Hubbard projectors and
the definition of the subspace projection operators.
In order to provide an unbiased analysis of different subspace
definitions, therefore, we employed the same
experimental lattice constant for all calculations.
In order to obviate intervention
in the construction of the correlated subspaces,
so far as possible, we carried out the DFT+$U$
calculations in the projector self-consistent 
formalism described in Ref.~\onlinecite{PhysRevB.82.081102}.
We also include, for the purposes of comparison, 
the results of conventional 
DFT+$U$ calculations using hydrogenic
$3d$-orbital Hubbard projectors 
(in which case there is no ambiguity in the 
representation for a given choice of projectors) 
which were used as the initial guess
for the projector self-consistency cycle~\cite{charges}.

In the projector self-consistent DFT+$U$
scheme of Ref.~\onlinecite{PhysRevB.82.081102}, the set of 
five converged NGWFs of 
maximal $3d$-orbital character on a transition-metal 
atom responsible for strong correlation effects
are selected as Hubbard projectors
to redefine the DFT+$U$ occupancy matrices for the total
energy minimization in the next projector iteration.
The energy is not directly minimized with respect to the
expansion coefficients of the Hubbard projectors
(since it would violate the variational principle
if either the Hubbard
projectors or $U$ were allowed to change
during energy minimization~\cite{eschrig}),
but the projectors are updated
in a manner reminiscent of  
the density-mixing method for solving non-linear 
systems~\cite{321305,*1965},
converging towards those
which equal a subset of the NGWFs
which minimize 
the DFT+$U$
energy functional which they themselves define.
The projector-update process  
alternates between direct
variational minimization of the total energy and
projector renewal until both are individually converged.

\subsection{Occupancies and magnetic 
dipole moments}

\begin{figure}
\includegraphics*[width=8.5cm]
{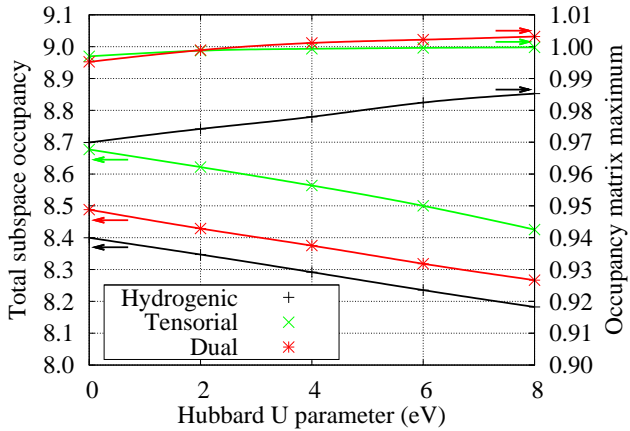}
\caption{(Color online) The total occupancy of a correlated
subspace in NiO (left axis) and the 
maximum element on the diagonal of the occupancy
matrix (right axis), as a function
of the interaction $U$. Values are computed with
orthonormal hydrogenic Hubbard projectors and self-consistent
NGWF projectors in both the ``dual'' and ``tensorial''
representations.}
\label{Fig:IMAGENiO-hydro_total_occ_vs_U}
\end{figure}

In agreement with a number of previous 
studies~\cite{PhysRevB.57.1505, 
PhysRevB.62.16392,0953-8984-20-32-325205}, 
we find that the LDA correctly favours 
antiferromagnetic ordering in NiO, albeit with diminished 
local magnetic moments and a greatly 
underestimated Kohn-Sham gap.
The DFT+$U$ correction enhances
the antiferromagnetic order with increasing $U$,
monotonically increasing the magnetic dipole moments. 
Also in accordance with previous 
work~\cite{PhysRevB.73.045110,
0953-8984-20-32-325205}, 
we have found that the
DFT+$U$ occupancy matrix and local
magnetic dipole moment associated with the 
correlated subspaces depend significantly
on the definition of the 
correlated subspace projection operator.

Turning first to the total occupancy of the 
correlated subspaces, shown in 
Fig.~\ref{Fig:IMAGENiO-hydro_total_occ_vs_U},
we find a steady decrease with increasing $U$
parameter, which is almost entirely due to the 
DFT+$U$ correction introducing 
a repulsive potential to the  
less-than-half occupied nickel $3d-e_g$
orbitals of the minority spin channel.
Conversely, we notice that for the largest 
element on the diagonal of the occupancy
matrix (which is almost identical to that of the other
orbitals of the same symmetry), 
DFT+$U$ introduces an
attractive potential that tends to fully occupy 
the corresponding orbital.

The maximal occupancy element for hydrogenic projectors, 
for those projectors most commonly used in DFT+$U$
which are not adapted to their chemical environment and
so cannot fully account for densities deviating from spherical 
symmetry, slowly approaches unity and 
we conjecture that a rather 
excessive $U$ value would be needed
to complete the orbital filling. On the other hand, if we look at 
self-consistent NGWF projectors in the ``dual'' 
representation, there is a tendency to overfill the 
most fully occupied Hubbard projectors, to wit
the occupancy begins to exceed unity beyond 
$U \approx 3~$eV. This latter affliction is a rather 
hazardous one for the DFT+$U$ functional, 
since the contribution to the energy correction
arising from orbitals exhibiting it may become negative
in severe cases;
this is incorrect behaviour for a penalty-functional in any case. 
The reason
behind this excessive occupancy is the spurious non-locality of the
Hubbard projector duals in the ``dual" representation, they
may pick up density contributions from all across the simulation cell.
On the contrary, when self-consistent projectors 
are used in the ``tensorial" representation, the
maximal matrix elements tend asymptotically to unity
with increasing $U$, as expected (reaching $0.9998$
at $U=8~$eV). 

\begin{figure}
\includegraphics*[width=8.5cm]
{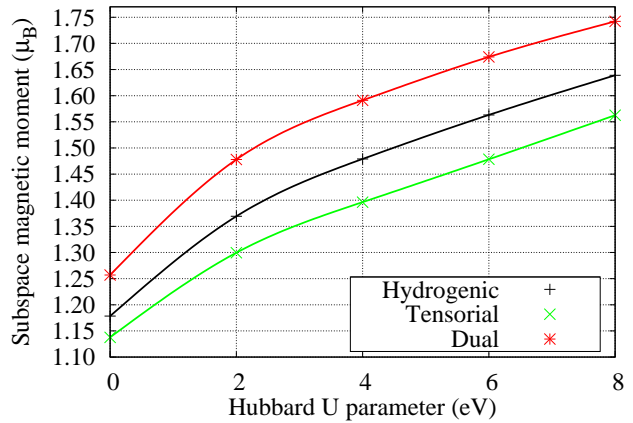}
\caption{(Color online) The projection of the magnetic dipole 
moment onto the DFT+$U$ correlated 
subspace on nickel atoms 
in NiO, computed as a function of the interaction $U$.
Values are computed as in 
Fig.~\ref{Fig:IMAGENiO-hydro_total_occ_vs_U}.}
\label{Fig:IMAGENiO-hydro_local_mag_vs_U}
\end{figure}

In order to test the dependence 
of our computed DFT properties
on the XC functional used for pseudopotential
generation, these dependencies known 
to be potentially substantial 
when non-linear core corrections are 
used~\cite{PhysRevB.57.2134}, 
we also performed our ``hydrogenic" calculations using
LDA~\cite{PhysRevB.23.5048} pseudopotentials with parameters closely matching to the GGA set~\cite{opium}. 
Moving from the latter to the former pseudopotentials, 
we observed a reduction of the local magnetic 
moment by $0.007~\mu_B$ at $U=0~$eV, up to $0.02~\mu_B$ at $U=4~$eV, whereat the reduction 
remains with further increase in $U$.
The total correlated subspace occupancy is rigidly increased by
approximately $0.02~e$. The maximum occupancy matrix 
changes by no more than $0.001$ electrons any $U$ tested.
The occupancy matrices thus depend
on the choice of pseudopotential, as expected, 
as do derived properties, 
but not sufficiently to influence our observed trends, indeed
by a small amount compared to the dependence on
the $U$ parameter and subspace projection definition.

Considering the local magnetic moment on the
nickel atoms, depicted in 
Fig.~\ref{Fig:IMAGENiO-hydro_local_mag_vs_U}
and defined as the difference of
the traces of the DFT+$U$ occupancy matrices of the 
two spin-channels,
we observe the expected increase with the $U$
parameter as the majority and minority channels of
the magnetization-carrying orbitals become increasingly
filled or emptied, respectively. 
The NGWF projectors in the ``dual" representation yield
 greater local magnetic moments than the 
 representation-independent hydrogenic projectors
and, in turn, those are larger than the moments
in the ``tensorial" representation.
Consequently, we would expect the exchange splitting
 which makes up a large contribution to the
insulating gap in this material (it is well-described within
unrestricted Hartree-Fock theory~\cite{PhysRevB.50.5041}), 
to follow the same trend. 
While this behaviour
may seem a somewhat unfavourable reflection on
the ``tensorial" representation, it is fully in line 
with our understanding that the ``dual" 
representation (or any related
delocalized ``Mulliken''-type analysis)
picks up additional contributions from 
magnetization densities of
neighbouring atoms  by construction. 
The previously demonstrated
strong dependence 
of the moments on the definition
of the subspace occupancy 
matrices~\cite{PhysRevB.73.045110},
taking the $4~$eV spread 
of $U$ values which approximately yield $1.48~\mu_B$ 
in our calculations as an example, 
demonstrates the 
hazard incurred by comparing
$U$ parameters used with DFT+$U$ methods that
differ in their technical details.

\subsection{Kohn-Sham eigenspectra}

\begin{figure}
\includegraphics*[width=8.5cm]
{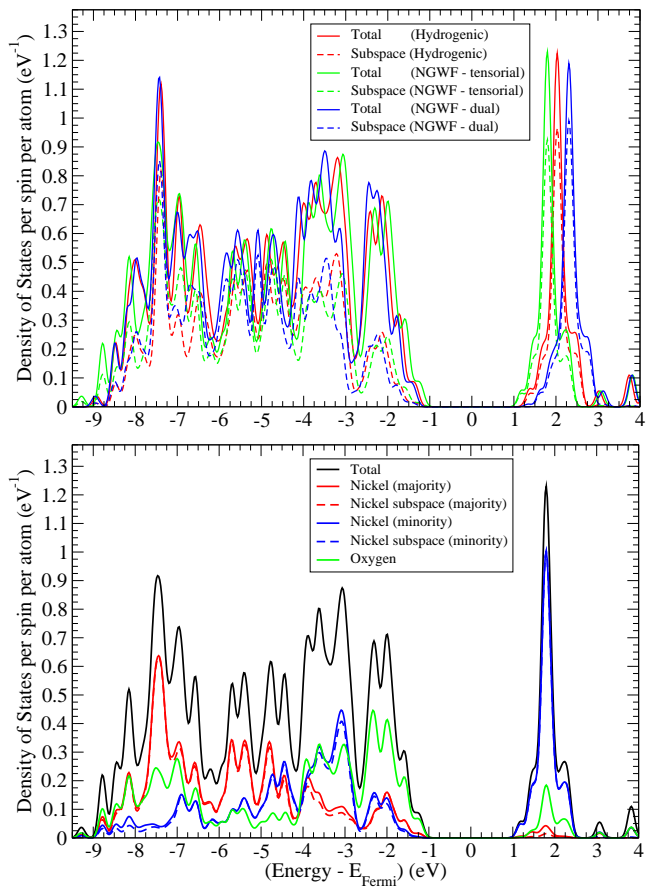}
\caption{(Color online) Top: Density of Kohn-Sham states 
per spin per atom of NiO
 at LDA+$U=6~$eV, together
with its projection onto the union of correlated
subspaces using hydrogenic Hubbard projectors and 
NGWF projectors in the  ``tensorial'' and ``dual''
representations. Bottom: 
The decomposition, in the NGWF-``tensorial'' representation, of the 
density of states for a given spin channel into its contributions from 
NGWFs on nickel atoms with magnetization
aligned (majority) and anti-aligned (minority) spins, the correlated
subspace projections of each, and the contribution due to
NGWFs on oxygen atoms.}
\label{Fig:NiOUeq6PDOS_combined}
\end{figure}

The Kohn-Sham eigenspectrum computed
for NiO using DFT+$U$ with both 
our ``best guess" 
system-independent hydrogenic projectors~\cite{charges} and
self-consistently determined NGWF projectors
agree closely. 
Moreover, in agreement
with previous studies of the dependence
on the occupancy matrix definition when using nonorthogonal
Hubbard projectors~\cite{PhysRevB.73.045110,0953-8984-20-32-325205},
 the representation dependence of 
spectral features is rather subtle
and is considerably less significant than the dependence on the
$U$ parameter. That is not to say, however, that the
differences yielded may be guaranteed to be fully
recovered by a self-consistent 
determination or arbitrary variation
of the interaction $U$, since we observe
different dependences on this parameter for different
spectral peaks depending on the 
subspace representation.

Considering, for example, a Hubbard $U$ value within
the range of values known to give reasonable agreement
with experiment, namely $U=6~$eV, 
we have shown the total density of states (DoS),
and its correlated subspace projection, in the three
representations of interest in the top panel of
Fig.~\ref{Fig:NiOUeq6PDOS_combined}.
The bottom panel shows
the decomposition of the ``tensorial" DoS into its contributions
from oxygen atoms and both predominantly spin-aligned 
(majority) or spin-antialigned (minority) nickel atoms.
Although all of the dominant features are
shared between the eigenspectra of the 
various representations, 
there are some discrepancies which are worth noting.
Most notable is the trend for the insulating gap
to open slightly, predominantly at the minority $e_g$ peak
at $\approx 2~$eV, as we go from ``tensorial" NGWF 
($2.35~$eV) to 
hydrogenic ($2.60~$eV) to ``dual" 
NGWF representations ($2.68~$eV). 
We attribute this  
to changes in the exchange 
splitting provided by the enhancement 
of the magnetic 
moment, which follows the same trend, 
as can be seen in
Fig.~\ref{Fig:IMAGENiO-hydro_local_mag_vs_U}.
The localized character of the valence band edge is not
significantly representation-dependent.

\begin{figure}
\includegraphics*[width=8.5cm]
{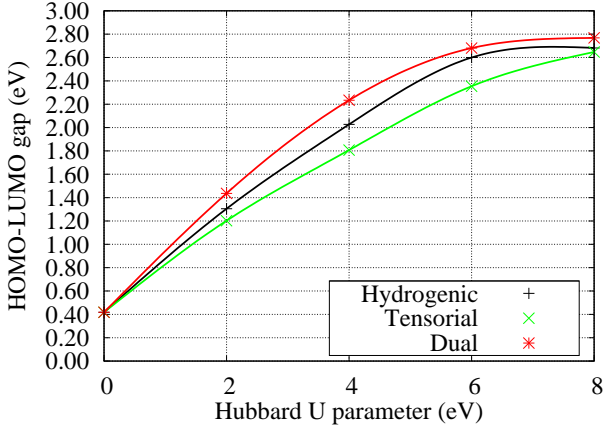}
\caption{(Color online) The Hubbard $U$ dependence
of the Kohn-Sham band-gap
of NiO
 at LDA+$U$. 
}
\label{Fig:IMAGENiO_gap_vs_U}
\end{figure}

We show the $U$-dependence
of the Kohn-Sham
insulating gap in the three DFT+$U$ correlated
subspace definitions in 
Fig.~\ref{Fig:IMAGENiO_gap_vs_U}.
In all cases, we recover the canonical DFT+$U$
description of this material.
With increasing interaction parameter
$U$ the tendency is that: the 
low-energy (primarily majority-channel $e_g$-like)
 peak falls deeper into the valence band as an 
 attractive potential is applied to fill it completely; 
 the strongly nickel $t_{2g}$-like valence-band edge
 at the LDA level gives way to hybridized oxygen $2p$
 character as the $t_{2g}$-like states are pushed to lower 
 energies; and the minority-channel Nickel $e_{g}$-like
 first peak in the conduction band is increased 
 in energy as its partial occupancy causes it to be 
 subjected to a repulsive corrective potential.
  
 Overall, we reiterate that the
 effects on the spectra due to the
 local or non-local construction of the
 Hubbard projector duals, at least in this material, 
 are not sufficiently great to 
 reasonably draw conclusions regarding the relative merit of methods 
 based on agreement, or otherwise, with 
 experimental observations. Rather, in this matter,
 points of principle such as the preservation 
 of tensorial invariance, or the
 avoidance of  occupancies exceeding unity
 (as observed in Fig.~\ref{Fig:IMAGENiO-hydro_total_occ_vs_U}),
 must therefore take precedence
 in our view.

\section{Copper phthalocyanine dimer}

Open-shell molecular systems containing transition
metal ions sometimes pose a 
challenge to first principles simulation within 
LDA-based approximations~\cite{AronJ.Cohen08082008}. 
This is partially due to the tendency of
such approximate XC functionals 
to excessively delocalize magnetization-carrying 
orbitals in such systems. As noted in 
Refs.~\onlinecite{PhysRevB.81.075403,
PhysRevB.65.184435,doi:10.1021/jp070549l,
PhysRevLett.97.103001}, 
both energetic properties, such as 
magnetic coupling,
and also spectroscopic features, such
as the nature of the insulating gap 
and multiplet splittings, can 
consequently be poorly reproduced by
such functionals.
Sophisticated \emph{ab initio} techniques such as the
$GW$ approximation and
local correlation methods such as DFT+$U$,
whose traditional realm of application 
lies in extended systems such as extended oxides and
their interfaces, are being increasingly applied to 
molecular systems and clusters (see for example
Refs.~\onlinecite{PhysRevB.80.155443,
PhysRevLett.97.103001,PhysRevB.65.184435,
doi:10.1021/jp070549l,
PhysRevB.81.075403}). 

It is thus of some importance, and perhaps timely, to consider
molecular systems on a similar 
footing to solids when considering the merit of projection
 methods for DFT+$U$. 
 The correlated orbitals in molecular systems may
be rather more spatially diffuse
and deviate further from spherical symmetry
than their counterparts in solids. 
As a result, the issue of Hubbard 
projector-dependence in DFT+$U$ and then the manner
in which the projection operator is constructed from those
projectors, particularly the degree of non-locality in the
Hubbard projector duals, can be expected to play a more
significant role in the description of molecular systems. 

With a view to analyzing the dependence on the
correlated subspace definition, or occupancy representation,
in the case of molecular systems, we applied
our methodology to the 
ground-state of a
binuclear open-shell (antiferromagnetically 
coupled) singlet complex, the 
copper phthalocyanine dimer
denoted $\alpha$-Cu({\small{\sc II}})Pc$_2$.
Crystalline CuPc is a semiconducting blue dye which, in
pure thin-film form and more exotic derivatives, 
is currently attracting
intense experimental and theoretical interest 
due to its potential for use as a flexible organometallic 
photovoltaic material~\cite{bao:3066}, as part of field-effect
transistors~\cite{peumans} and, due to its 
magnetic functionality, in spintronic
data storage or processing devices~\cite{Cinchetti}.
In this system, two correlated subspaces
delineated by copper $3d$-like states are spatially
well separated, with approximately $3.77$~\AA~between 
centres,
and there is minimal electronic 
bonding between the localized orbitals in the open Cu-$3d$ 
shells in the two 
approximately planar moeities. The result is a 
very weak indirect-exchange, i.e., acting via 
intermediary delocalized ligand states, $S=\frac{1}{2}$ 
antiferromagnet with a Heisenberg exchange 
coupling constant of $J \approx -1.50K$; for a detailed
analysis of this mechanism see 
Ref.~\onlinecite{PhysRevB.77.184403}.

It is important to emphasise that 
although corrective techniques for
localized correlation effects such as DFT+$U$ 
have been shown to be
somewhat beneficial in the 
context of organometallic molecules~\cite{springerlink:10.1007/s00339-008-5022-0,
0957-4484-18-42-424013},
the are by no means the only, 
or perhaps favourable, methods 
for such systems.
For Cu({\small{\sc II}})Pc$_2$, as we go on to show,
the magnetization-carrying copper $3d$ 
orbitals are partly delocalised
and thus not fully recovered by DFT+$U$. 
Systems of this type have been described with particular
success, notably~\cite{marom:164107,
PhysRevB.77.184403,evangelista:124709}, 
using hybrid XC functionals 
comprising a fraction of non-local Hartree-Fock 
exchange more appropriate to these molecules. 

The \textsc{ONETEP} 
method~\cite{onetep1,*ChemPhysLett.422.345}
was used, as before, 
with $\Gamma$-point Brillouin zone sampling,
norm-conserving pseudopotentials~\cite{PhysRevB.41.1227,opium}
and a set of nine NGWFs
($4s$, $4p$ and $3d$) for
copper ions, four each for carbon and nitrogen
($2s$ and $2p$) and one for hydrogen ($1s$).
 An NGWF cutoff radius of $5.3$~\AA~and 
 an equivalent kinetic-energy cutoff of 
 $1000~$eV was used.
 The spin-polarized 
 PBE~\cite{PhysRevLett.77.3865} generalized-gradient 
XC
 functional was employed.
An un-solvated and hydrogenated
gas-phase dimer model was extracted~\cite{nina} from the
 $\alpha \left( + \right)$Cu({\small{\sc II}})Pc$_2$ 
 polymorph structure, 
 with a stacking angle of $65.1$ degrees 
 and a distance
 between molecular planes of $3.42$~\AA, 
 giving a lateral offset of $1.58$~\AA, as
 reported from transmission electron diffraction analysis
 described in Ref.~\onlinecite{hoshino}.
 A simulation cell of 
 $30$\AA$ \times 30$\AA $\times 20$\AA~provided 
 an interatomic spacing
 between periodic images of at least $13.5$~\AA~in plane
 and $16.5$~\AA~out of plane.
 
 \subsection{Magnetic dipole moments}
 
  \begin{figure}
\includegraphics*[width=8.5cm]
{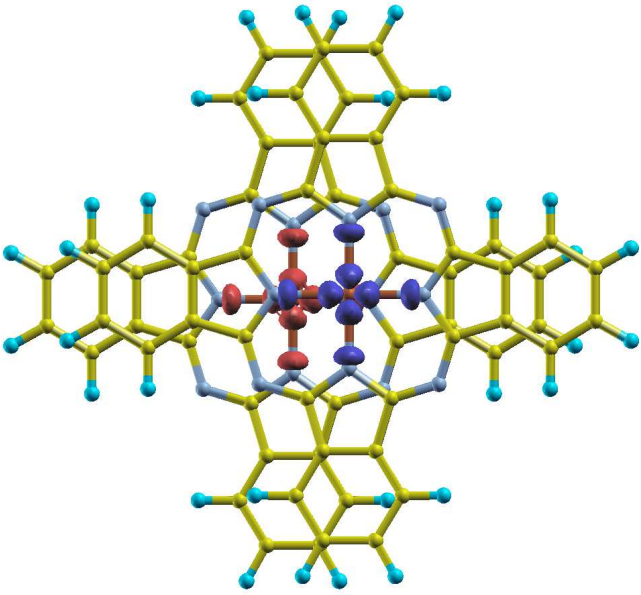}
\caption{(Color online) Spin-density 
isosurfaces at 5\% of maximum in
Cu({\small{\sc II}})Pc$_2$
at projector self-consistent 
GGA+$U=6~$eV within the NGWF-``tensorial'' 
subspace representation.}
\label{Fig:CuPc2_Ueq6_singlet_SCF1_fireball_spindensity}
\end{figure}

The open-shell singlet fragments of the 
Cu({\small{\sc II}})Pc$_2$ system consist of 
single spins, i.e., a moment of $1~\mu_B$
on each copper centre, aligned antiparallel
 with respect to each other. 
Since, however, approximate XC 
functionals may lower the energy by delocalizing and partially
occupying magnetization-carrying 
orbitals~\cite{AronJ.Cohen08082008}, a
diminished value for the local moment is often recovered in practice.
The DFT+$U$ method seeks to ameliorate
this condition in two complementary ways, that is by
introducing a derivative-discontinuity to the 
energy functional which penalises fractional occupancies of the
spin-orbitals defined by the subspace projections
and also, in doing so, 
by effectively constraining the Kohn-Sham spin-orbitals
to more closely resemble the (usually more localized) 
spatial form of the correlated subspace.

\begin{figure}
\includegraphics*[width=8.5cm]
{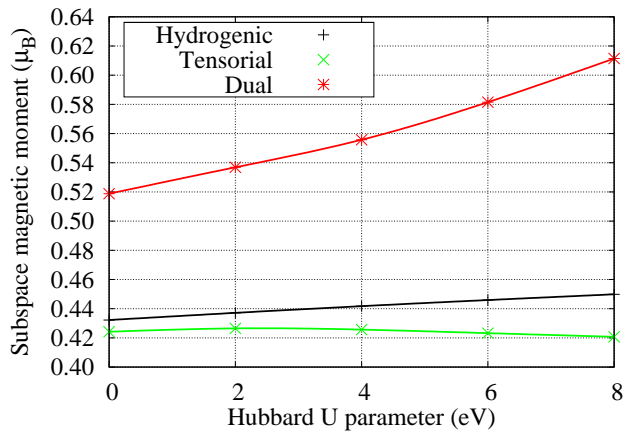}
\caption{(Color online) The average magnitude of the projection of the 
magnetic dipole
onto the correlated subspaces of Cu({\small{\sc II}})Pc$_2$, 
plotted as a function of $U$ for various definitions of the
 subspace projection.}
\label{Fig:IMAGE-singlet_hydro_local_mag_vs_U}
\end{figure}

In spite of this, the correlated subspace projected
magnetic dipole moments, shown in
Fig.~\ref{Fig:IMAGE-singlet_hydro_local_mag_vs_U},
indicate that the DFT+$U$ method does not effectively 
localize the magnetization density to the copper $3d$
manifold for any reasonable value of the $U$ parameter.
Using conventional hydrogenic Hubbard projectors, with our
best guess for the radial profile~\cite{charges}, 
we see that there is only a
very slight increase in the local moment with $U$.
Switching to self-consistent 
projectors in the ``tensorial" representation, 
we find that the moment is
effectively $U$-independent and
reduced with respect to the hydrogenic result.

Conversely, the ``dual"
representation yields a greater magnetic moment than the
 ``tensorial" representation, by approximately 
$0.1~\mu_B$ at $U=0~$eV, increasing steadily at 
 a rate of $\approx 0.02~\mu_B eV^{-1}$. 
 The reason for this discrepancy, and failure of DFT+$U$
 in this regard, is 
 understood via the atom-decomposed
 Mulliken analysis of the magnetization density, which gives
 $0.10-0.12~\mu_B$ on each nitrogen atom which 
 is a nearest-neighbour to copper, 
 irrespective of either the representation or
 the $U$ parameter. 
 Notwithstanding their adaptation to the 
 molecular environment, the self-consistent NGWF projectors 
 remain predominantly on the copper ion and 
 do not have sufficient weight on the neighbouring 
in-plane nitrogen $2p$ orbitals to capture the 
 magnetization density associated with them. As a result, 
 in the same manner as the 
 conventional projectors, they 
 fail to retrieve the magnetization to the copper 
 $3d_{x^2 - y^2}$ orbital within DFT+$U$. 
 The ``dual" representation, however,
partially  overcomes this obstacle,
 due to the dual Hubbard projectors extending over all of the
 delocalized states in the system
 including those that contribute to the magnetization density.

\subsection{Kohn-Sham eigenstates}

The accepted 
understanding~\cite{marom:164107,evangelista:124709,
PhysRevB.77.184403} 
of the spectroscopic 
nature of the gap in the copper phthalocyanine monomer is
that the HOMO level is dominated by a doubly-occupied
$a_{1u}$ orbital which consists of a superposition
of Carbon $p_z$ orbitals delocalized on the pyrrole rings
of both monomer units, 
while the spectroscopically correct LUMO level is 
also a delocalized doubly-degenerate orbital, 
of $e_g$ symmetry composed of a superposition of
$\pi$ orbitals on pairs of macrocycle Carbon atoms. 
We may expect some minor differences in the spectroscopic properties
in the dimer system with respect to the monomer,  
due to $\sigma$-bonding between moieties,  
but for the main features to be preserved.

\begin{figure}
\includegraphics*[width=8.5cm]
{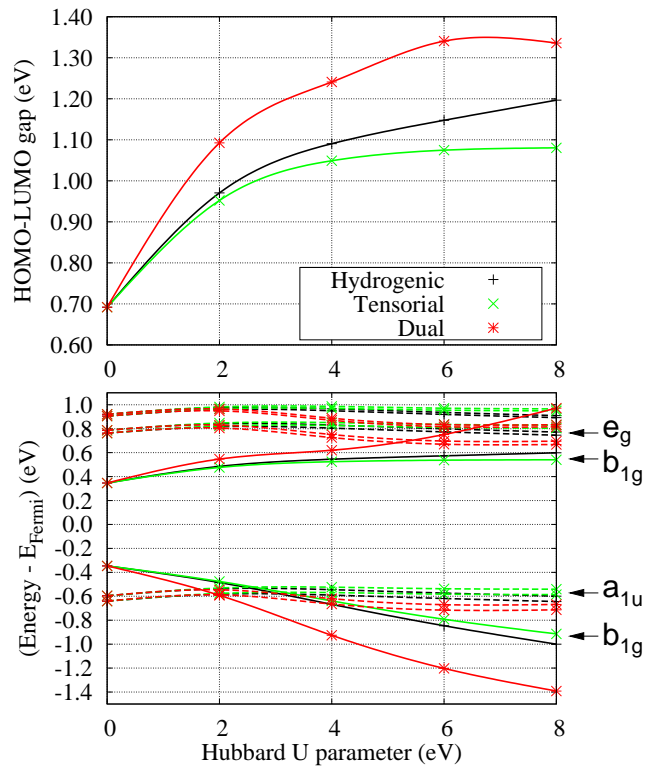}
\caption{(Color online) The HOMO-LUMO
energy gap (top) and the
energy levels adjacent to the Fermi energy (bottom)
of Cu({\small{\sc II}})Pc$_2$,
 plotted as a function of $U$. 
 In the bottom panel, solid lines show energy levels 
 of states of predominantly Cu-centered
 $b_{1g}$ character, and 
 to which the 
 DFT+$U$ correction strongly applies, 
 while dashed lines show energy levels of 
 states of more delocalized
 nature.}
\label{Fig:IMAGE-singlet_hydro_gap_bands_vs_U}
\end{figure}

It has previously been shown that, due to self-interaction errors,
LDA and GGA-type XC functionals do not
correctly reproduce the qualitative ordering of states
close to the Fermi level in the 
monomer~\cite{marom:164107,
springerlink:10.1007/s00339-008-5022-0}.
The DFT+$U$ insulating gap of the dimer system, 
within various representations, is shown in 
Fig.~\ref{Fig:IMAGE-singlet_hydro_gap_bands_vs_U}, along
with the $U$-dependence of the states nearest the Fermi 
energy. 
For the spin-polarized PBE functional, we find a 
gap of $0.7~$eV for the dimer, whose nature is a
charge-transfer excitation between $b_{1g}$ orbitals 
on either moiety. The $b_{1g}$ orbital is that which carries
the magnetization density in the dimer, consisting primarily of 
copper $3d_{x^2 - y^2}$ $\sigma$-bonded 
to in-plane nitrogen $2p$. 
The representation dependence of the HOMO-LUMO 
gap follows the same trend as the local magnetic moment, 
due to the DFT+$U$ correction 
to the Coulomb-repulsion
gap being somewhat augmented by an 
associated enhancement to the 
exchange splitting.  
In the case of the HOMO
orbital, a small value of $U$ is needed to push the 
singly-occupied $b_{1g}$ state to its spectroscopically 
correct position below the $a_{1u}$ state and the effect
is rather more strongly pronounced in the ``dual" representation
than in the spatially-localized methods. 

The incomparability between 
Kohn-Sham eigenenergies
with either experimental optical or photoemission spectra
notwithstanding, it 
is perhaps worth noting some similarities and
differences between
our computed Kohn-Sham levels for the
dimer system and the gas-phase
ultraviolet photoelectron spectra (UPS) and
x-ray absorption near-edge structure (XANES) reported in 
Ref.~\onlinecite{evangelista:124709}. Turning first
to the valence band edge, the UPS spectra confirm that
single-molecule CuPc possesses 
a doubly-occupied HOMO of 
pyrrole-delocalised $a_{1u}$
character, while a further weak 
ionisation peak at $800~$meV above that
is consistent with
the magnetisation-carrying copper $3d_{x^2 - y^2}$-based
$b_{1g}$ orbital seen, albeit substantially
closer to the valence band edge for moderate $U$
values, in 
Fig.~\ref{Fig:IMAGE-singlet_hydro_gap_bands_vs_U}.
This may suggest a suppression of spin-splitting between
the $b_{1g}$ levels in the dimer over the monomer system, 
or may be due to an underestimation of 
experimental transition energies 
which is not sufficiently alleviated by DFT+$U$. In the case of
the lowest conduction bands, carbon K edge XANES 
spectra assign a large pyrrole carbon character to the 
LUMO, consistent with a delocalised and degenerate 
$e_g$ type orbital.
Due to the differing localisation regions of the carbon $1s$
and singly-occupied $b_{1g}$ orbital, as noted 
in Ref.~\onlinecite{evangelista:124709}, 
such excitations are neglected and so we cannot 
compare our prediction that the latter orbital 
lies somewhat below the $e_g$ level 
for all except the ``dual" representation at high $U$ values.
We agree on the proximity of these latter levels
with previous monomer calculations using the PBE 
functional~\cite{marom:164107,0957-4484-18-42-424013}.

The ``tensorial" and
``hydrogenic" representations have similar effects, as expected;
the effect of projector self-consistency is rather small 
in this system. In the case of the virtual orbitals, the 
localized $b_{1g}$ character of the LUMO
persists for the ``tensorial" and 
``hydrogenic" methods, which agree quite closely, while 
$U \geq 6~$eV is sufficient to expose a delocalized $e_g$
orbital as LUMO in the ``dual" representation. 
There is necessarily some small perturbative
effect on delocalized orbitals induced by changes to those 
which are DFT+$U$ corrected, which is evident in all projection
techniques but noticeably stronger in the ``dual" representation. 

The overall result is that for this hybridized correlated system, 
the ``dual" representation recovers
the expected magnetic dipole moment 
with significantly more success than the 
fully localized projections. The spectroscopic 
nature of the insulating gap is also 
recovered to a greater degree for a given value of $U$.
We would contend, however, that it does so for 
reasons not expected in 
the DFT+$U$ method. Specifically, where
the local magnetic moment as measured by the ``dual"
projectors increases with increasing $U$, the spatial distribution
of this increase is made up both of the region immediately 
surrounding the copper ion and spatially diffuse contributions, as 
opposed to the ``tensorial" or conventional orthonormal ``hydrogenic" 
contributions, with which we are guaranteed to include only
subspace-localized densities. 

\section{Concluding remarks}

We have presented a revised
 formalism for the construction of
projection operators, and consequently the occupancy matrices,
of strongly-correlated subspaces 
using nonorthogonal Hubbard
projector functions in \emph{ab initio}
methods such as DFT+$U$ and DFT+DMFT.
In contrast to the previously proposed 
``full"~\cite{PhysRevB.58.1201}, 
``on-site"~\cite{eschrig} and 
``dual"~\cite{PhysRevB.73.045110} representations, 
our ``tensorial" definition preserves the
important property of tensorial
invariance in the total occupancy of each subspace, 
the total energy and the ionic forces, by construction.
The required expressions for the 
tensorially-invariant DFT+$U$ energy functional and the
resulting potential and ionic forces, have been presented.

Localized nonorthogonal basis functions for Kohn-Sham 
states are frequently 
used to represent the Hubbard projectors, in practice,
either for reasons of computational convenience or to achieve
projector self-consistency~\cite{PhysRevB.82.081102}.
We have shown, however, that 
it is inappropriate to continue to 
identify the dual space (and the metric tensor) of
the basis functions with the dual space of Hubbard 
projectors on each site.
For molecular systems, in particular, the unexpected 
discrepancy with respect to orthonormal projectors that is
thereby introduced may be significant.
The resulting projector duals (contravariant vectors)
are unsuited to constructing a correction 
for localized correlation effects,
generally being delocalized across the entire simulation cell. 
When using delocalized projector duals, 
moreover, a tensor-incompatible 
symmetrization of the projection 
operator is needed to ensure a Hermitian potential.
This may result in unphysical occupancy matrix elements
and an uncontrolled  
action of the corrective potential which it defines.
Put simply, additional non-local corrections
are introduced in the ``dual" representation 
which are extraneous to the
requirement of accounting for the nonorthogonality of the
Hubbard projectors.

Our tensorial formalism may be implemented in any
methodology which makes use of
 a nonorthogonal set of functions
to define each correlated subspace. Since it inherently
preserves the spatial localization of 
Hubbard projector duals,  
it is also less computationally expensive and simpler to 
implement in linear-scaling methods, in practice, than
the ``on-site" or ``dual" representations which employ
delocalized dual projectors.
In order to alleviate the remaining arbitrariness in 
DFT+$U$ and related methods in the nonorthogonal case, 
the tensorial formalism may be combined with both
 a projector self-consistency algorithm~\cite{PhysRevB.82.081102} 
 or any one of a number
 of available first-principles methods for the $U$
  parameter~\cite{PhysRevB.58.1201,
PhysRevB.71.035105,PhysRevLett.97.103001,
PhysRevB.39.1708,*PhysRevB.43.7570,
*PhysRevB.74.235113,PhysRevB.70.195104,*PhysRevB.81.245113,PhysRevB.74.125106}; the latter remains as an avenue
for future investigation.
 
 It is our hope that we have dispelled 
 some of the ambiguities surrounding this topic
 which we feel have arisen inevitably
 as a result of the neglect of the invaluable tensor notation.
 As the use of linear-scaling \emph{ab initio}
 approaches becomes increasingly widespread, we 
 envisage that this work may aid the routine 
 implementation of sophisticated functionality in 
 the nonorthogonal bases,
 obviating the expenditure of explicit orthonormalization.
 
\begin{acknowledgments}

We would like to thank Nicholas Hine
for helpful discussions. 
This research was supported by EPSRC, 
RCUK and the National University of Ireland.
Calculations were performed on the Cambridge HPCS 
Darwin computer
under EPSRC grant EP/F032773/1 and the
UK National Supercomputing Service HECToR computer
with support from the UKCP consortium.

\end{acknowledgments}

\appendix

\section{Orthonormal Hubbard projectors}

Orthonormal sets of Hubbard projectors, as well 
as nonorthogonal sets, may provide a compact and
accurate representation of the correlated subspaces
and we would not wish to detract 
from their value and ease of use.
In the orthonormal case, the Hubbard projectors 
equal their own duals with respect to their subspace, 
and the metric tensors reduce to  
Kronecker delta functions.

 If one performs an inverse
  L\"{o}wdin transform~\cite{lowdin} from
 an orthonormal set of projectors to a nonorthogonal frame
  using the matrix square root of covariant and contravariant
  metrics on a particular 
  correlated subspace, $O^{\frac{1}{2}}$ 
and  $O^{-\frac{1}{2} }$,  
respectively, then
   the pre-multiplicative scalar $U$ parameter for that site 
   remains identically the same, since for each site 
   (if $n$ and $n'$ index orthonormal projectors and
    $m$ and $m'$ index their nonorthogonal counterparts) we have, 
    supposing $n, n', m, m' \in\{1,\ldots,M^{(I)} \}$,
  \begin{align}   
{}& \sum_{n n'} U_{n n' n n'}  \nonumber
= \sum_{n n' n'' n'''} U_{n n' n'' n'''} 
\delta_{n n''}  \delta_{n' n'''} \\ \nonumber
={}& \sum_{n n'} \sum_{ n'' n'''} 
U_{n n' n'' n'''}  
\sum_{m m'}  O^{\frac{1}{2}}_{ n m}
  O^{ -\frac{1}{2}   m n''} 
   O^{\frac{1}{2} }_{ n' m'}  O^{ -\frac{1}{2}  m' n'''} \\
  ={}& \sum_{m m'} U_{m m'}^{ \;\;\;\;\; m m'} \equiv M^{(I)2} U .
  \nonumber
 \end{align}  
 
  Thus, when the Coulomb 
  interaction is approximated by a
  pre-multiplicative scalar $U$ times the identity, 
  we retain its usual 
  interpretation  
  as the averaged screened Coulomb repulsion 
  between densities in the subspace
  described by the Hubbard projectors,
  regardless of whether or not
  the Hubbard projectors are orthonormal.  
  
  \section{Invariance under  generalized L\"{o}wdin transforms}
  
  As suggested in 
  Ref.~\onlinecite{0953-8984-20-32-325205},
   generalized definitions of the L\"{o}wdin transform
  may be envisaged whereby the metric
  tensor is raised to an arbitrary power $A$, 
  as is its inverse, and the canonical
  L\"{o}wdin transform $A= \frac{1}{2}$ has the
  status of a special case.
  Since, however, by construction
  \begin{align} 
\delta_{n }^{\;\; n'}
=  
 \sum_{m \in \mathcal{C}^{(I)}}O^{(I)(A)}_{ n m} 
  O^{(I) (-A)  m n'} \nonumber 
  =   
   \sum_{\gamma \in \mathcal{S} }S^{(A)}_{ n \gamma}
  S^{ (-A)  \gamma n'},
  \end{align} 
the fully averaged scalar $U$ is invariant
  under such transformations, independent of the 
  exponent $A$,
  regardless of whether the subspace metric tensor 
  $O_{\bullet \bullet}$ or, 
  in the ``dual'' representation case, the metric 
  $S_{\bullet \bullet}$ on the space
  spanned by all basis functions, $\mathcal{S}$, is used.
     
In the latter case of $S_{\bullet \bullet}$, 
  the generalized L\"{o}wdin 
  transformation exponent $A$ varies the nonorthogonality of 
  the representation of the 
  occupancy matrices or, equivalently, 
  (since the basis set metric $S$
  introduces spurious contributions to the 
  occupancy matrix from across
  the simulation cell) the degree of non-locality of the correction. 
 The dependence of computed ground-state 
 properties and of the Kohn-Sham 
 gap of a variety of materials on $A$, as reported in 
 Ref.~\onlinecite{0953-8984-20-32-325205},
 demonstrates, in our view, the ambiguity of 
 population analysis measures,
 and hence corrections such as DFT+$U$, which
 are built from tensorially inconsistent (necessarily symmetrized)
  occupancy matrices where
  delocalized
 Hubbard projector duals of the form 
 $\lvert \varphi^{(I)m} \rangle = 
\sum_{\gamma \in \mathcal{S} } 
\lvert \varphi^{(I)}_{\gamma} 
\rangle  S^{ (-A)  \gamma m} 
 $ are used in the construction of the Hubbard projectors.
 
The observation of Ref.~\onlinecite{0953-8984-20-32-325205}
 that the $A$ parameter bears
influence on computed properties accords well with
our arguments on the unsuitability of the metric 
$S^{\bullet \bullet}$ (that for the 
basis functions in the entire simulation cell) in constructing
localized self-interaction corrections such as DFT+$U$, 
since that parameter effectively
controls the superfluous spatial 
delocalization of the Hubbard projector
duals and hence
 the severity of the tensorial inconsistency in
the DFT+$U$ functional. 
By varying $A$, the occupancy matrix for the ``dual''
representation subject to a generalized L\"{o}wdin transformation,
given by
\begin{equation} \nonumber
\sum_{\gamma, \delta \in \mathcal{S} }
S^{ (A)  m \gamma}
\langle \varphi^{(I)}_\gamma \rvert \hat{\rho} \lvert 
\varphi^{(I)}_\delta \rangle S^{ (1-A)  \delta m'},
\end{equation}
picks up differing non-local contributions (densities
from outside the correlated subspace). 
Spurious non-local contributions
are incorporated 
for all values of $A$, moreover.

On the contrary, in
the ``tensorial" representation, the generalized L\"{o}wdin 
transformed occupancy matrix
\begin{equation} \nonumber
\sum_{m'', m''' \in  \mathcal{C}^{(I)}} 
O^{(I) (A) m m''}
\langle \varphi^{(I)}_{m''} \rvert \hat{\rho} \lvert 
\varphi^{(I)}_{m'''} \rangle O^{(I) (1-A) m''' m'},
\end{equation}
contains no contributions from outside the correlated subspace
it is describing, for any value of $A$. Both the trace of this matrix
and the trace of its square are entirely independent of $A$, since 
$O^{(I) (1-A) m''' m} O^{(I) (A) m m''} = O^{(I) m''' m''}$.
Thus, by construction, the DFT+$U$ correction is  invariant under
generalized L\"{o}wdin transformations
and so is unambiguously defined, for a given
choice of projectors, when the 
appropriate subspace-local metric 
tensor $O^{\bullet \bullet}$ is used 
to build the projection operator.


\begin{thebibliography}{72}
\bibitem [{cha()}]{charges}%
  \BibitemOpen
  \href@noop {} {}\bibinfo {note} {The effective nuclear charge $Z$ used to
  construct the hydrogenic projectors, which may influence predicted
  observables in the case of system-independent Hubbard
  projectors~\cite{PhysRevB.82.081102,PhysRevB.58.1201} but which does not
  significantly influence results at projector self-consistency, was estimated
  by fitting the hydrogenic radial probability density to that of the
  corresponding valence pseudo-orbitals in the sense of least squared
  deviations (resulting in $Z=9.02$ and $Z=9.10$ for $3d$-orbital projectors in
  nickel and copper, respectively).}\BibitemShut {Stop}%
\end{thebibliography}
\end{document}